\documentclass[12pt]{iopart}
\usepackage{iopams}
\pdfoutput=1
\usepackage{amstext}
\usepackage{amssymb} 
\usepackage{latexsym}
\usepackage[english]{babel}
\usepackage[dvips]{graphicx}

\usepackage{graphicx}
\usepackage{dcolumn}
\usepackage{ulem}
\usepackage{color}
\usepackage[colorlinks=true,urlcolor=blue,citecolor=blue, linkcolor = blue]{hyperref}
\usepackage{epsfig}
\usepackage{epsf}

\begin{document}

\title[The XYZ chain with Dzyaloshinsky-Moriya interactions]
{The XYZ chain with Dzyaloshinsky-Moriya interactions: from spin-orbit-coupled lattice bosons 
to interacting Kitaev chains}

\author{Sebastiano Peotta$^{(1)}$, Leonardo Mazza$^{(2)}$, Ettore Vicari$^{(3)}$, Marco Polini$^{(4)}$, Rosario Fazio$^{(2,5)}$, and Davide Rossini$^{(2)}$} 

\address{(1) Department of Physics, University of California, San Diego, La Jolla, CA 92093, USA}
\address{(2) NEST, Scuola Normale Superiore and Istituto Nanoscienze-CNR, I-56126 Pisa, Italy}
\address{(3) Dipartimento di Fisica dell'Universit\`a di Pisa and INFN, Largo Pontecorvo 3, I-56127 Pisa, Italy}
\address{(4) NEST, Istituto Nanoscienze-CNR and Scuola Normale Superiore, I-56126 Pisa}
\address{(5) Center for Quantum Technologies, National University of Singapore, Singapore}

\begin{abstract}
Using the density-matrix renormalization-group algorithm (DMRG) and a finite-size scaling analysis, 
we study the properties of the one-dimensional completely-anisotropic spin-1/2 XYZ model with Dzyaloshinsky-Moriya (DM) interactions. 
The model shows a rich phase diagram: depending on the value of the coupling constants, the system can display different kinds of ferromagnetic order and Luttinger-liquid behavior. Transitions from ferromagnetic 
to Luttinger-liquid phases are first order. We thoroughly discuss the transition between different ferromagnetic phases, which, in the absence of DM interactions, belongs to the XX universality class. 
We provide evidence that the DM exchange term 
turns out to split this critical line 
into two separated Ising-like transitions and that
in between a disordered phase may appear.
Our study sheds light on the general problem 
of strongly-interacting spin-orbit-coupled bosonic gases
trapped in an optical lattice and 
can be used to characterize the topological properties of superconducting nanowires in the presence of an imposed
supercurrent and of interactions.
\end{abstract}



\maketitle

\section{Introduction}
\label{introduction}

Ultracold atoms in optical lattices constitute a unique tool to study equilibrium
as well as non-equilibrium properties of many-body 
quantum systems. 
The versatility of these setups, offered by the possibility of
manipulating and initializing them in a wide range of regimes 
for several choices of atomic species, has 
lead to an impressive number of 
breakthroughs in the study of strongly correlated systems of 
bosons and fermions, as well as of their mixtures~\cite{Lewenstein:2007,Bloch:2008}.  
By dressing atomic states with properly-designed laser fields 
it is possible to engineer synthetic gauge 
fields~\cite{Dalibard:2011, Galitski_2013}, thus paving the way for 
the exploration of Bose-Einstein condensates (BEC) and degenerate Fermi gases in presence of external magnetic fields~\cite{Spielman_2009_SOA}
and spin-orbit coupling~\cite{Spielman_2011_SOA, Zwierlein_2012_SOA, Zhang_2012_SOA}, 
even in the presence of optical lattices~\cite{Bloch_2013_SOA, Bloch_2013b_SOA, Ketterle_2013_SOA, Ketterle_2013b_SOA, Bloch_2014_SOA}.

In particular, the experimental realization 
of a spin-orbit-coupled (SOC) BEC~\cite{Spielman_2011_SOA} 
has brought to the attention of the community the problem of investigating 
the interplay between interactions and non-Abelian gauge fields. In the Abelian case (i.e.~for an external magnetic field), this interplay 
leads to the spectacular physics of the fractional quantum Hall effect~\cite{Tsui_prl_1982}.
In the case of weak interactions,
the theoretical characterization has been
thorough and detailed~\cite{weaksoc}. 
However, ultracold bosonic atoms can  be driven 
into the strongly-interacting 
regime by means of an optical lattice, 
and for deep enough potentials a transition to a Mott insulating phase takes 
place~\cite{Lewenstein:2007,Bloch:2008}. 
Whereas the density distribution of the cold atom gas in a Mott insulating phase is constrained to yield 
an integer number of particles per site, multi-component bosonic gases can display a variety
of possible phases due to the underlying pseudo-spin degrees of freedom. For example, 
different types of ``magnetic'' orderings, both in the insulating and superfluid regimes, can occur~\cite{Lewenstein:2007}. 

So far only two- and three-dimensional lattice systems 
have been investigated (see for example~\cite{Grass:2011,Cole:2012,
Radic:2012,Cai_2012,Mandal:2012,Wong:2012,Qian:2013,Zhang:2013, Grass_2012, Grass_2013} and references therein)
and the phase diagram has been shown to feature several 
intriguing properties. 
The superfluid phase
can display exotic features and it can be  spatially  modulated, whereas
in the Mott insulator (MI) phase the bosonic Hamiltonian 
can be mapped~\cite{Cole:2012} onto an XYZ-model with 
Dzyaloshinsky-Moriya (DM) interactions~\cite{Dzyaloshinsky:1958,Moriya:1960}. 
The phase diagram of this model in one spatial dimension (1D) has not been completely mapped out till now. 
In this Article we address this problem by means of a
Density-Matrix Renormalization-Group (DMRG) algorithm~\cite{Schollwock:2011, Dummies}. 
The main results of this analysis are presented in Fig.~\ref{fig:phase_diagram_XYZ-DM}. The implications 
of these results on the magnetic phases of SOC bosonic MIs are discussed.

Remarkably, this study sheds light also on the topological 
properties of 1D nanowires~\cite{review1,review2}. 
As first pointed out by Kitaev~\cite{Kitaev:2001}, 1D fermionic systems
undergo a topological phase transition in the presence of p-wave
pairing. The topological phase is characterized
by the presence of zero-energy Majorana modes localized at the end points of the chain.
Using our results, we are able to discuss the robustness  
of such edge modes to the simultaneous presence of interactions and of an 
external magnetic field, which couples to the fermionic
motional degrees of freedom. 
This study widens previous analysis on 
interacting Kitaev wires~\cite{Gangadharaiah:2011, 
Stoudenmire:2012, Lobos_2012, Braunecker_2013}. 

Our Article is organized as follows. 
In  Section~\ref{model} we introduce our model, i.e.~the 
spin-1/2 Hamiltonian of the XYZ model with DM interactions. We 
highlight its connections to the mentioned bosonic and fermionic models.
The main DMRG results concerning the characterization of the phase diagram
are reported in Section~\ref{diagram}
and are supplemented by the appropriate finite-size scaling analysis. 
In Section~\ref{sec:SOCbosons} we discuss these results
from the point of view of
lattice bosons and spinless fermions mentioned above.
We conclude our work with Section~\ref{conclusions}, 
where a summary of our results is presented together 
with an outlook on future investigations.

\section{The Model}
\label{model}

We study the XYZ spin-1/2 Hamiltonian
with a DM interaction term ($\hbar = 1$)~\cite{Dzyaloshinsky:1958,Moriya:1960}:
\begin{equation}
 	\hat{\cal H}  = \hat{\cal H}_{\perp} + \hat{\cal H}_z  \, ,
\label{ham}
\end{equation}
where
\begin{eqnarray}
\label{hamperp}
 	\hat{\cal H}_{\perp}  &=& - \sum_j  \left(J e^{i \varphi} 
 	\hat S^+_j   \hat S^-_{j+1}+
 	J_{\Delta}  \hat S^+_j \hat S^+_{j+1}  \right)
	+ \mbox{H.c.}~,\\
 	\hat{\cal H}_z &=&  J_z  \sum_j \hat S^z_j  \hat S^z_{j+1}~.
\label{hamz}
\end{eqnarray}
Here $J>0$ and $\hat S^{\alpha}_j$ ($\alpha = x,y,z$) are spin-$1/2$ operators 
on the $j$-th site ($\hat S^{\pm}_j$ are the corresponding raising/lowering 
operators). The Hamiltonian contains short-range interactions 
characterized by three coupling constants: $J e^{-i\varphi}$, $J_{\Delta}$, and $J_z$.
Because of the term controlled by $J_{\Delta}$, 
which is here taken to be a real number,
the phase $\varphi$ cannot be gauged away even in an open chain
and is related to a DM interaction. 
Indeed, by expressing Hamiltonian~(\ref{hamperp}) in terms of  
$\hat S^{x}_i$ and $\hat S^{y}_i$,  one gets
$
 	\hat{\cal H}_{\perp} = 
 	- \sum_i \left( J_x \hat S^x_{i} \hat S^x_{i+1} + 
 	J_y   \hat S^y_{i} \hat S^y_{i+1} +
 	D \hat {\bf z}  \cdot \hat {\bf S}_{i}  \times \hat {\bf S}_{i+1} \right)  
$
with the identification  $ J e^{i \varphi} = (J_x+J_y+i 2 D)/4$ and $J_{\Delta} = (J_x - J_y)/4$. 
In the rest of the Article we  discuss the 
zero-temperature phase diagram of the Hamiltonian~(\ref{ham}) 
using  the parametrization given in Eqs.~(\ref{hamperp}) and~(\ref{hamz}).

\subsection{Related Models: Spin-Orbit-Coupled Lattice Bosons and Fermionic Nanowires}
\label{subsec:SOCbosons}

As anticipated in the Introduction, the model defined in Eq.~(\ref{ham}) 
is related to two paradigmatic cold-atom and condensed-matter models.
It is useful at this stage to make these mappings explicit, 
although already known in the literature,
so that our findings can be compared more easily with related bibliography.

The Hamiltonian~(\ref{ham}) represents an effective model for 
a lattice system loaded with two bosonic species 
(i.e. a hyperfine doublet in the context of ultracold atoms) 
with an anisotropic interaction and spin-orbit coupling. 
The corresponding 1D Bose-Hubbard (BH) Hamiltonian reads:
\begin{equation}
\hspace*{-1cm}
     \hat{\cal H}_{\rm BH} = 
    \sum_j \left[-t\left( \hat b^\dagger_j e^{i\alpha \tau^y} \hat b_{j+1} \! + \! \mbox{H.c.} \! \right)
      + \! \frac{g_1}{2} \left(\hat n_{j} \right)^2 \! + \! \frac{g_2}{2}\left( \sum_{\beta, \gamma}
       \hat b^{\dagger}_{j,\beta} 
      \, \tau^z_{\beta, \gamma} \, \hat b_{j, \gamma}
      \right)^2\right] \! .
\label{discrete_hamiltonian1}
\end{equation}
Here $ \hat b_{j} = (\hat b_{j,\uparrow},\hat{  b}_{j,\downarrow})$ is a bosonic annihilation  operator for the two components at site $j$, which are for brevity addressed with the pseudo-spin $\{\uparrow, \downarrow \}$ notation; 
$ \hat{n}_j$ is the on-site density operator 
and $\tau^\beta$ are the Pauli matrices which act on the pseudo-spin degrees of freedom 
($\tau^z_{\beta,\gamma}$ denotes the matrix elements of $\tau^z$).
The first term in Eq.~(\ref{discrete_hamiltonian1}) represents the hopping, whose amplitude 
is $t$; the angle $\alpha \neq 2 \pi m$,  $m \in \mathbb Z$, quantifies the strength 
of  spin-orbit coupling (in the continuum limit the 
momentum operator would couple to the $y$-component  of the spin). 
The last two terms describe interactions between bosons: the term 
proportional to $g_1$ is the standard BH repulsive term, while 
the  one controlled by $g_2$ fixes a preferred orientation in spin space. 

Note that we have chosen two orthogonal preferred directions for the spin-orbit coupling 
and interaction anisotropy, thereby fully breaking the $\text{SU}(2)$ spin symmetry. 
The choice of a spin-orbit axis aligned along $z$
produces a less interesting model, as
the corresponding spin-orbit term can be gauged away in an open chain.  
If $ g_1 \gg | g_2|$ and one is well inside the MI phase, only spin degrees of 
freedom play a role. In this limit it is therefore convenient to introduce an effective spin Hamiltonian. 
A straightforward second-order expansion in the small parameter $t/g_1$ yields a model 
which is formally equivalent to the one in Eq.~(\ref{ham}),
modulo a different labeling of the axes. Introducing the shorthand $g \equiv g_2 / g_1$, the parameters
of the two models are related by the following identities:
\begin{eqnarray}
	J_z &= & -\frac{4 t^2}{g_1} \, \frac{1}{1- g} \, , \nonumber \\
	J e^{i \varphi} &= & \frac{4 t^2}{g_1} \, \frac{1}{1- g} \,
	\frac{1-g}{2(1+ g)} e^{i 2 \alpha} \, , \nonumber\\
	J_{\Delta} & = & - \frac{4 t^2}{g_1} \, \frac{\cos (2 \alpha)}{1- g} \,
	\frac{g}{1+g}~. 	\label{eq:BH:SOC}
\end{eqnarray}
The most relevant effect of spin-orbit coupling is to introduce a DM interaction~\cite{Cole:2012, Radic:2012}. Thus, the phase diagram that we are going to present
is relevant for future experiments with synthetic gauge fields in 1D optical lattices loaded with two bosonic species.

Interestingly, studying the Hamiltonian~(\ref{ham}) is  also important  
for the problem of interacting topological insulators and,
more specifically, for the robustness of zero-energy Majorana 
modes in semiconducting nanowires~\cite{review1,review2,Kitaev:2001}.
By means of a Jordan-Wigner transformation, the Hamiltonian~(\ref{ham}) can be mapped
onto a 1D model of interacting spinless fermions 
with hopping amplitude $Je^{-i\varphi }$, p-wave pairing potential $J_{\Delta}$ 
and a nearest-neighbor interaction $J_z$:
\begin{equation}
  	\hat\mathcal{  H}_{\rm K}  =  \sum_j \!\bigg[ \, - \, (Je^{i\varphi}\hat{c}_j^\dagger
  	\hat{c}_{j+1}
   	+J_{\Delta} \hat{c}_j\hat{c}_{j+1}+ \mbox{H.c.})  \; + \;
	 J_z~\bigg(\hat{m}_j -  \frac 12 \bigg)
	 \bigg(\hat{m}_{j+1} - \frac 12 \bigg) \!\bigg].
\label{p-wave}
\end{equation}
Here $\hat c_j$ annihilates a fermion at site $j$ and $\hat m_j \equiv \hat c^\dagger_j \hat c_j $ is the usual 
density operator.
The complex phase $\varphi$ represents the coupling to an external magnetic
field, which induces a finite supercurrent into the system.
The interplay of this term with 
nearest-neighbor interactions has not been fully investigated yet.

\subsection{Exactly Solvable Cases}
\label{subsec:exact}

In this Section we present some properties of the XYZ-model~(\ref{ham}) that hold for special properties of the microscopic couplings,
where an exact solution is available.

\subsubsection{$\varphi = 0$ and $\varphi = \pi$. --- }

It is useful to recall what happens to the XYZ-model~(\ref{ham}) 
 when $\varphi = 0$. In this case an exact solution is known~\cite{Baxter_book, XYZ_phase_diagram}. 
In the thermodynamic limit the system spontaneously breaks the $\mathbb Z_2$ symmetry along the axis 
with the largest value of $|J_\alpha|$ ($\alpha = x,y,z$). 
For $J_\alpha > 0$ there is ferromagnetic order, while 
$J_\alpha < 0$ yields antiferromagnetic (N\'eel) order. 
The system is critical whenever there are two couplings that are equal and their absolute value exceeds that of the third one; in that case a Luttinger liquid (LL) phase appears.
Considering the ($J_{\Delta}/J, J_z/J$) plane, in the spirit of the parametrization of  Eq.~(\ref{hamperp}), the XYZ model is thus critical for $J_{\Delta} = 0$, $|J_z| \leq 2J$ 
(the equality corresponds to the ferromagnetic and antiferromagnetic Heisenberg models), 
and for $J_{\Delta}  = \pm (|J_z|/2- J) $ for $|J_{z}| \geq 2J$.

The case $\varphi = \pi$ is completely equivalent, since a unitary rotation
connects the model for $\{J, J_{\Delta}, J_z, \varphi=\pi\}$
with $\{J, -J_{\Delta}, J_z, \varphi=0\}$.

\subsubsection{$\varphi \neq 0, \pi$; $J_z = 0$. --- }

The most relevant exactly solvable case for 
$\varphi \neq 0, \pi$ is the case $J_z=0$.
As highlighted by Eq.~(\ref{p-wave}), the model~(\ref{ham}) can be mapped  into
a free fermion model and is thus exactly solvable~\cite{Kontorovich, Siskens1, Siskens2}.
The system is gapless for $|J_{\Delta}/J| \leq |\sin \varphi|$.
The most important effect of the DM interaction is thus to extend the 
critical line appearing for $J_{\Delta} = 0$ to a  region of  finite width.

\subsubsection{$\varphi \neq 0, \pi$; $J_\Delta = 0$. --- }

Interestingly, also the case $J_{\Delta} = 0$ yields an exactly solvable model~\cite{Perk_1976}.
In this case a unitary transformation can be used to gauge away the quantity $\varphi$, so that 
Eq.~(\ref{ham}) reduces to the well-known XXZ model. 
The system is gapless for $|J_z/J| \leq 2$ and displays 
power-law decaying correlations, which are typical of a LL.
However, because of the rotation needed to gauge away $\varphi$,
correlations are twisted into the $x$-$y$ plane. 

\subsubsection{Symmetries. --- }

When $\varphi \neq 0$ the number of symmetries of the system is relatively
small; nonetheless there are a few ones which yield 
important information.
\begin{enumerate}
 \item Rotation of $\pi/2$ in the $x-y$ plane: 
 $\hat S_j^x \to \hat S_j^y $;
 $\hat S_j^y \to -\hat S_j^x $;
 $\hat S_j^z \to \hat S_j^z $. 
 This unitary transformation changes $J_{\Delta} \to - J_{\Delta}$, leaving the other coupling constants unchanged. 
 The sign of $J_\Delta$ is therefore unessential.
 \item Inversion with respect to the center of the chain:
 $\hat S_j^\alpha \to \hat S_{L-j}^\alpha $.
 This unitary transformation changes $J e^{- i \varphi} \to  J e^{i \varphi}$, 
 leaving the other coupling constants unchanged. 
 The sign of $\varphi$ is therefore unessential.
 \item Rotation of $\pi$ in the $x$-$y$ plane, only at even sites~\cite{Korepin}:
 $\hat S_j^x \to -\hat S_j^x $;
 $\hat S_j^y \to -\hat S_j^y $;
 $\hat S_j^z \to \hat S_j^z $ for $j$ even. 
 This unitary transformation changes $J e^{- i \varphi} \to  J e^{-i (\varphi+ \pi)}$ 
 and $J_{\Delta} \to - J_\Delta$, leaving the other coupling constants unchanged. 
 Together with the previous symmetry, it implies that the phase $\varphi$ can be taken in the interval $[0, \pi/2]$.
\end{enumerate}

\section{Phase diagram}
\label{diagram}

We now present the main results of this Article, i.e.~the 
zero-temperature phase diagram of the XYZ-model 
with DM interactions~(\ref{ham}), 
obtained via a DMRG study.  
Our numerical simulations were performed for systems with open boundary
conditions up to $L = 400$ sites, 
keeping at most $m = 120$ states. 
We checked that the location of the phase boundaries is not  
affected by the value of the cut-off $m$.  
We focus only on the ferromagnetic region of the phase diagram 
centered around $ J_{z}/J \sim -2$
because this is the most relevant case for bosons in optical lattices (see Section~\ref{subsec:SOCbosons}). 
Without loss of generality, we take $J_{\Delta}/J \geq 0$. 

\begin{figure}
\begin{center}
\includegraphics[width=0.85\columnwidth]{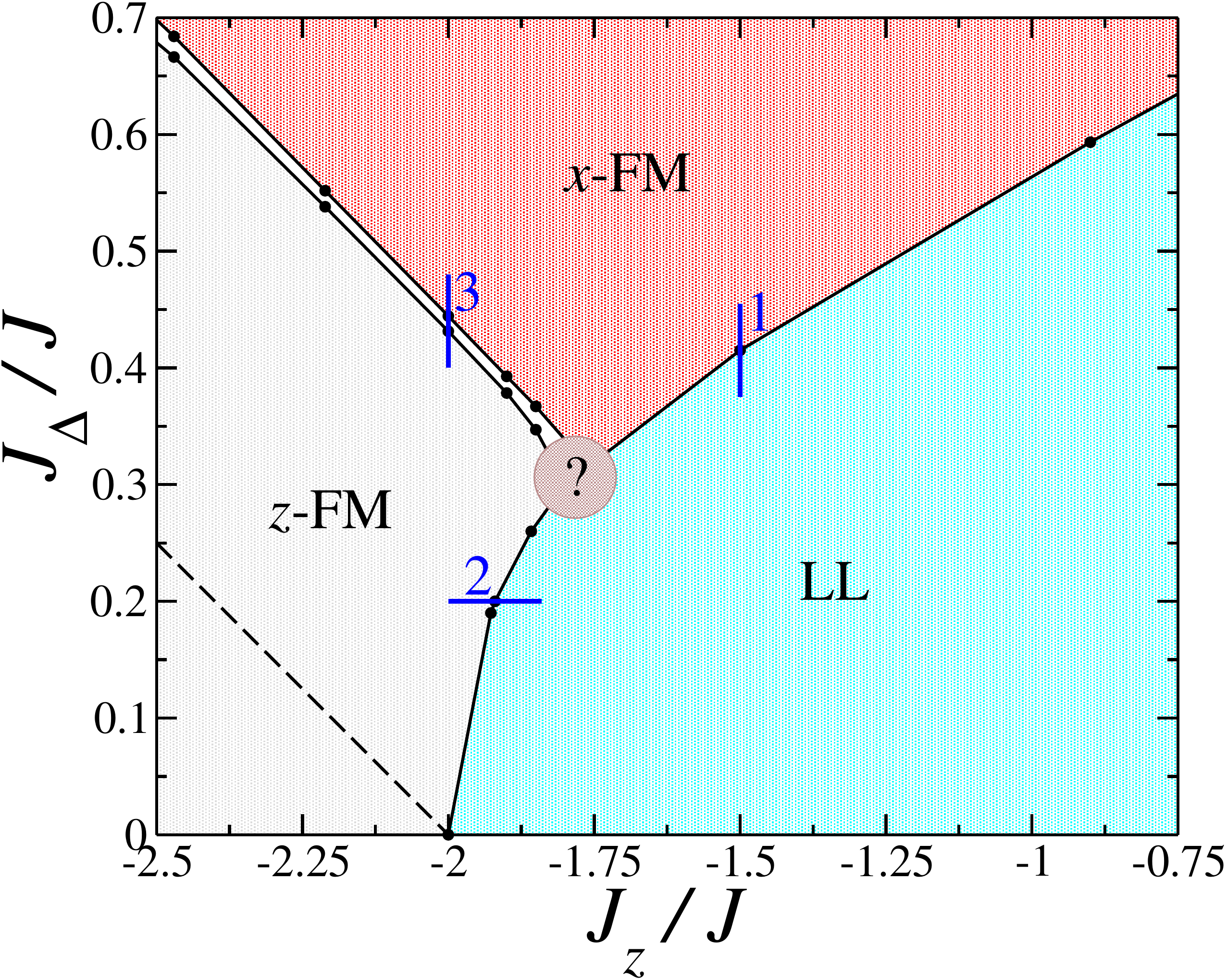}
\end{center}
\caption{(color online). 
  Zero-temperature phase diagram of the XYZ model with DM interaction
  as defined by Hamiltonian~(\ref{ham}) for $\varphi = 1$.
  Different colors denote the various phases:  The critical Luttinger liquid (LL) phase is depicted in blue,
  while the two ferromagnets ($x$-FM and $z$-FM) are in red and 
  in gray, respectively.
  The circle with a question mark identifies a region 
  which could not be reliably investigated
  and that may host a triple point.
  The dashed line denotes the transition between the two ferromagnetic phases occurring at $\varphi = 0$. The analysis of our numerical data supports 
  the existence of an intermediate disordered (white) region 
  separating the two ferromagnetic ones.
  Straight blue segments indicate the three cuts along which the various phase transitions are specifically addressed in the text.}
\label{fig:phase_diagram_XYZ-DM}
\end{figure}

\begin{figure}
\begin{center}
  \includegraphics[width=0.6\columnwidth]{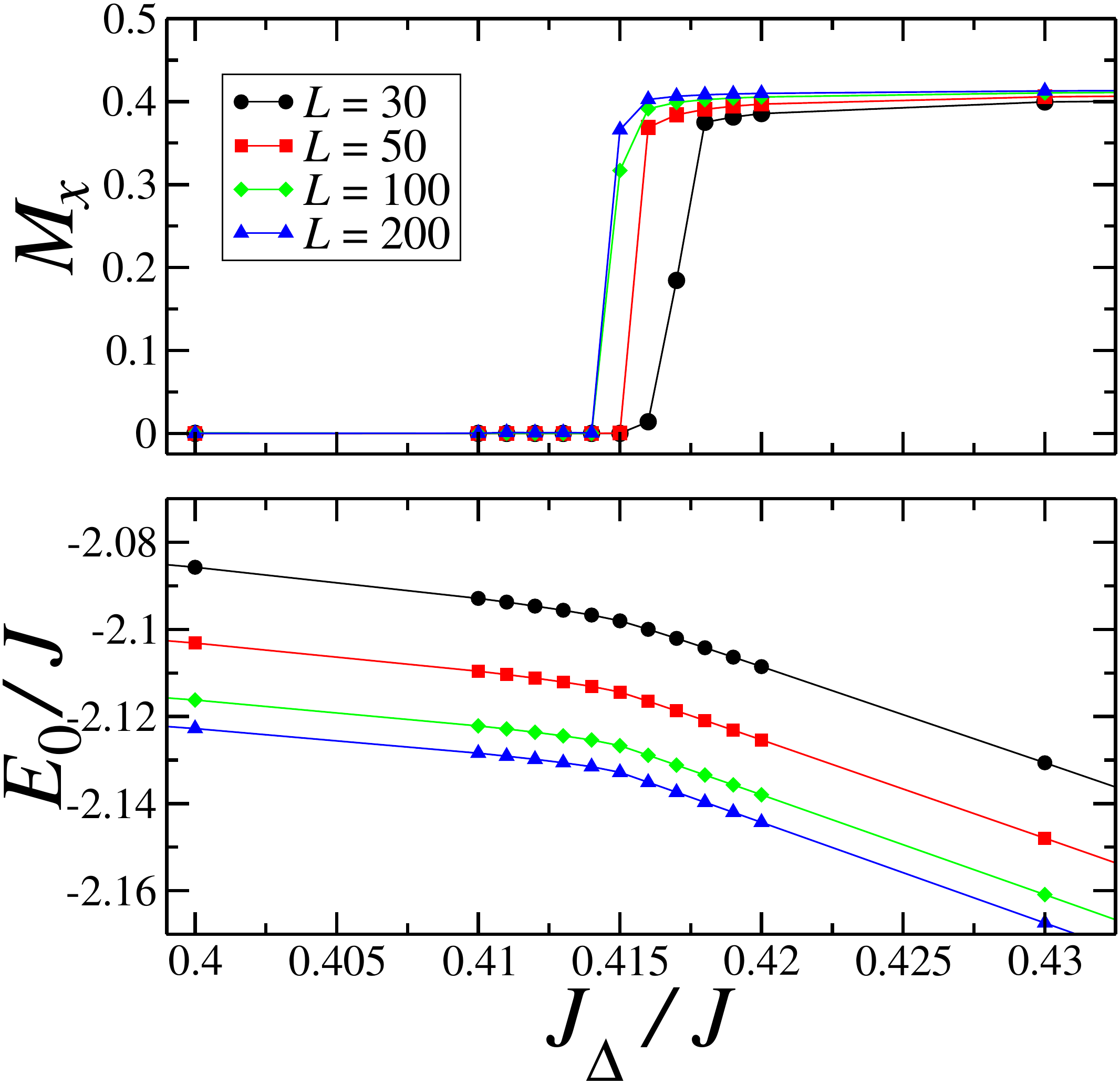}
\end{center}
  \caption{(color online).
    Study of the LL - ferromagnet transition along cut ``$1$'' 
    ($J_z/J = -1.5$) in the phase diagram in Fig.~\ref{fig:phase_diagram_XYZ-DM}.
    Upper panel: magnetization along $x$ as a function of $J_\Delta/J$.
    Different symbols and colors denote data for various system sizes.
    A discontinuity in $M_x$ signaling a first-order transition
    is observed at $(J_{\Delta}/J)_c \sim 0.415$.
    Lower panel: ground-state energy (in units of $j$ and per site) as a function of $J_\Delta/J$.
    The discontinuity in its first derivative locates 
    the transition point and the result is in agreement with the magnetization data.}
  \label{fig:LL-xFM} 
\end{figure}

The phase diagram of~(\ref{ham}) 
is shown in Fig.~\ref{fig:phase_diagram_XYZ-DM} for $\varphi = 1$. 
The study of other values of $\varphi$ resulted only in quantitative 
differences. 
The most relevant quantum phases are those which characterize the model also for $\varphi = 0$, 
namely, two ferromagnetic phases with different orientation (along the $z$ and $x$ axes) 
and a LL region.
As already discussed, the DM interaction 
is responsible for the finite width of the gapless region;  
furthermore, it rigidly shifts the transition between 
the two ferromagnetic phases 
(the case $\varphi=0$ is plotted 
in Fig.~\ref{fig:phase_diagram_XYZ-DM} with a dashed line).
One of the most interesting features of this phase diagram is the absence of
a direct transition between the ferromagnetic phases, 
where a new intermediate disordered region appears
(white area in Fig.~\ref{fig:phase_diagram_XYZ-DM}). 
As we will discuss below,
the LL-to-ferromagnet transitions are first order, 
whereas according to our analysis
a self-consistent description in terms of two
Ising-like critical lines can be formulated for
the transition between the ferromagnetic phases.
All the different phases seem to converge
in a region that could not be reliably analyzed 
because of accuracy problems 
while dealing with sizes $L>400$ (see~\ref{app:error}); 
the existence or absence of a triple point could not be assessed.
We now analyze the different transitions in more details; 
data will be presented for parameters running along the blue segments in Fig.~\ref{fig:phase_diagram_XYZ-DM}. 

\subsection{Phase Transition between a Ferromagnetic Phase and a Luttinger-Liquid}

\begin{figure}
\begin{center}
  \includegraphics[width=0.6\columnwidth]{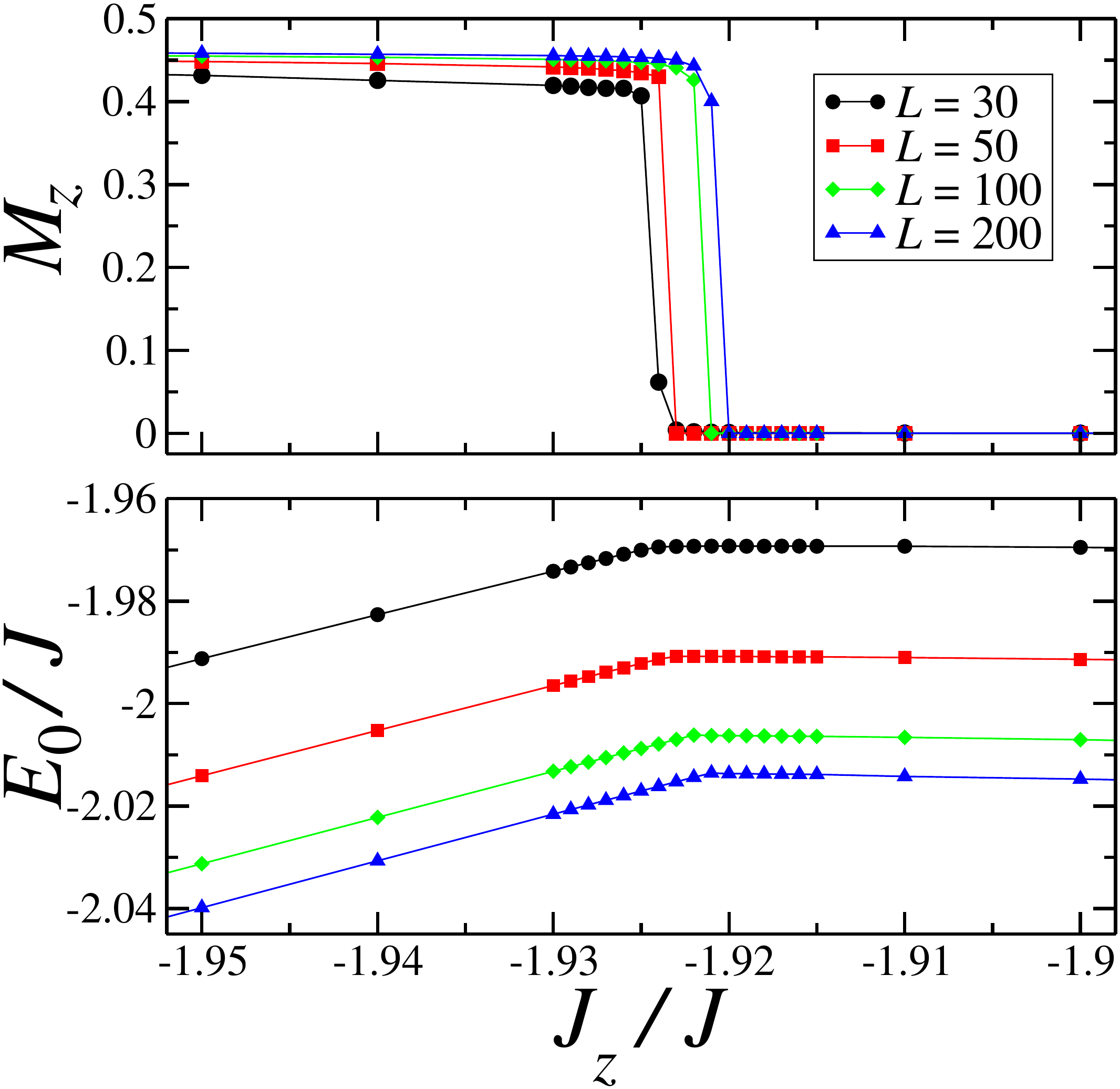}
\end{center}
  \caption{(color online).
    Study of the LL - ferromagnet transition along cut ``$2$''  ($J_\Delta/J = 0.2$)
    in the phase diagram in Fig.~\ref{fig:phase_diagram_XYZ-DM}.     
    Upper panel: magnetization along $z$ as a function of $J_z/J$.
    A discontinuity in $M_z$ signaling a first-order transition
    is observed at $(J_{z}/J)_c \sim -1.922$.
    Lower panel: ground-state energy (in units of $J$ and per site) as a function of $J_z/J$.
    The discontinuity in its first derivative locates 
    the transition point and is in agreement with the magnetization data.}
  \label{fig:LL-zFM}
\end{figure}

Let us first consider the transition between the LL and 
the ferromagnetic phases, cuts ``$1$'' and ``$2$''.
Ferromagnetically ordered phases can be distinguished either by measuring the magnetization $M_\alpha \equiv  \sum_i \langle   
\hat S^\alpha_i \rangle /L$ ($\alpha = x,y,z$), or by analyzing the asymptotic behavior of correlation functions.  
We show results based on the former indicator: 
no significant advantages were 
noticed by computing correlation functions.
For a finite chain of length $L$, 
spontaneous symmetry breaking is forbidden: the two lowest-energy states are non-magnetic 
and their degeneracy decreases exponentially with system size.
However, already
for $L \sim 100$ the degeneracy is far too small to be resolved by DMRG simulations.
To avoid numerical complications,
in the study of the ferromagnetic -- LL
phase boundary we
break the symmetry by adding two small magnetic fields $\mu_{\rm B} B_{\rm edge} \approx 10^{-5} J$ 
acting on the spins at the end points of the chain. 
(We have checked that our results do not depend on the value of such fields). 
In Figs.~\ref{fig:LL-xFM} and~\ref{fig:LL-zFM} we show the magnetization of the system across the cuts ``1'' and ``2'', respectively. 
Our data clearly display ferromagnetic order as a function of $J_{\Delta}$ or as a function of  $J_{z}$. 
A finite magnetization appears above a critical value of $J_{\Delta}$ ($x$-direction) 
and below a critical value of $J_{z}$ ($z$-direction). 
The phase transition between the LL phase and any of the two 
ferromagnetic phases is of the first order: 
this is signaled by a discontinuity both in the magnetization and in the first 
derivative of the ground-state energy.

Regarding the gapless phase, we can certainly conclude that no ferromagnetic order is present, even if
we did not make detailed simulations to analyze its properties. 
The blue region in Fig.~\ref{fig:phase_diagram_XYZ-DM} can be confidently classified as an extension of the gapless LL phase of
the XXZ model (at $J_\Delta = 0$), 
as it can be shown in a perturbative approach with respect to $J_{\Delta}$.

\subsection{Phase Transition between Two Ferromagnetic Phases}

We now investigate the phase diagram between the two ferromagnetic
phases, denoted by $x$-FM and $z$-FM in
Fig.~\ref{fig:phase_diagram_XYZ-DM}.  In the pure XYZ chain, the
disappearance of one kind of order (for instance along $z$) coincides
with the appearance of another kind of order (for instance along $x$).
It is a continuous phase transition of the XX universality class,
akin to that present in the XY chain, which is enforced by the
symmetries of the model. 

Such symmetries are lost in the presence of a
DM interactions, i.e., when $\varphi\ne 0$.
In order to investigate their effects we present a finite-size scaling
of the ratios~\cite{PV-14} $R_{\alpha} = \xi_{\alpha}/L$, where
$\xi_{\alpha}$ is the correlation length for the $\alpha=x,z$
component of the spin variables, i.e.,
\begin{equation}
  	\xi_{\alpha} = 
\sqrt{\frac{\sum_r r^2 \big\langle   \hat S_{i}^\alpha \,   \hat S_{i+r}^\alpha \big\rangle}
    	{2 \sum_{r} \big\langle  \hat S_{i}^\alpha \,  \hat S_{i+r}^\alpha \big\rangle}};
    	\qquad i = L/2 \; .
\label{eq:xixz}
\end{equation}
These quantities are particularly useful to identify continuous
transitions characterized by diverging length scales for the
correlation functions $\langle S_i^\alpha S_{i+r}^\alpha\rangle$.
Indeed around such critical points and for
large enough $L$, they are expected to behave 
as~\cite{Fisher:1973,Vicari:2002},
\begin{equation}
  R_\alpha = f \big( \delta_\alpha \cdot  L^{1/\nu} \big) + \ldots
\label{ralphasca}
\end{equation}
where $\delta_\alpha \equiv J_\Delta/J-(J_{\Delta}/J)_{c,\alpha}$
controls the distance from the critical point, and $\nu$ is the
length-scale critical exponent. The dots indicate scaling corrections
which are generally suppressed by powers of the inverse size~\cite{PV-14}.
Therefore, as implied by Eq.~(\ref{ralphasca}), 
the presence of a crossing point among data sets for different sizes $L$
provides the evidence of a critical
point. The slope at the crossing point is controlled by the universal
exponent $\nu$ associated with the universality class of the transition.

\begin{figure}
\includegraphics[width=0.5\columnwidth]{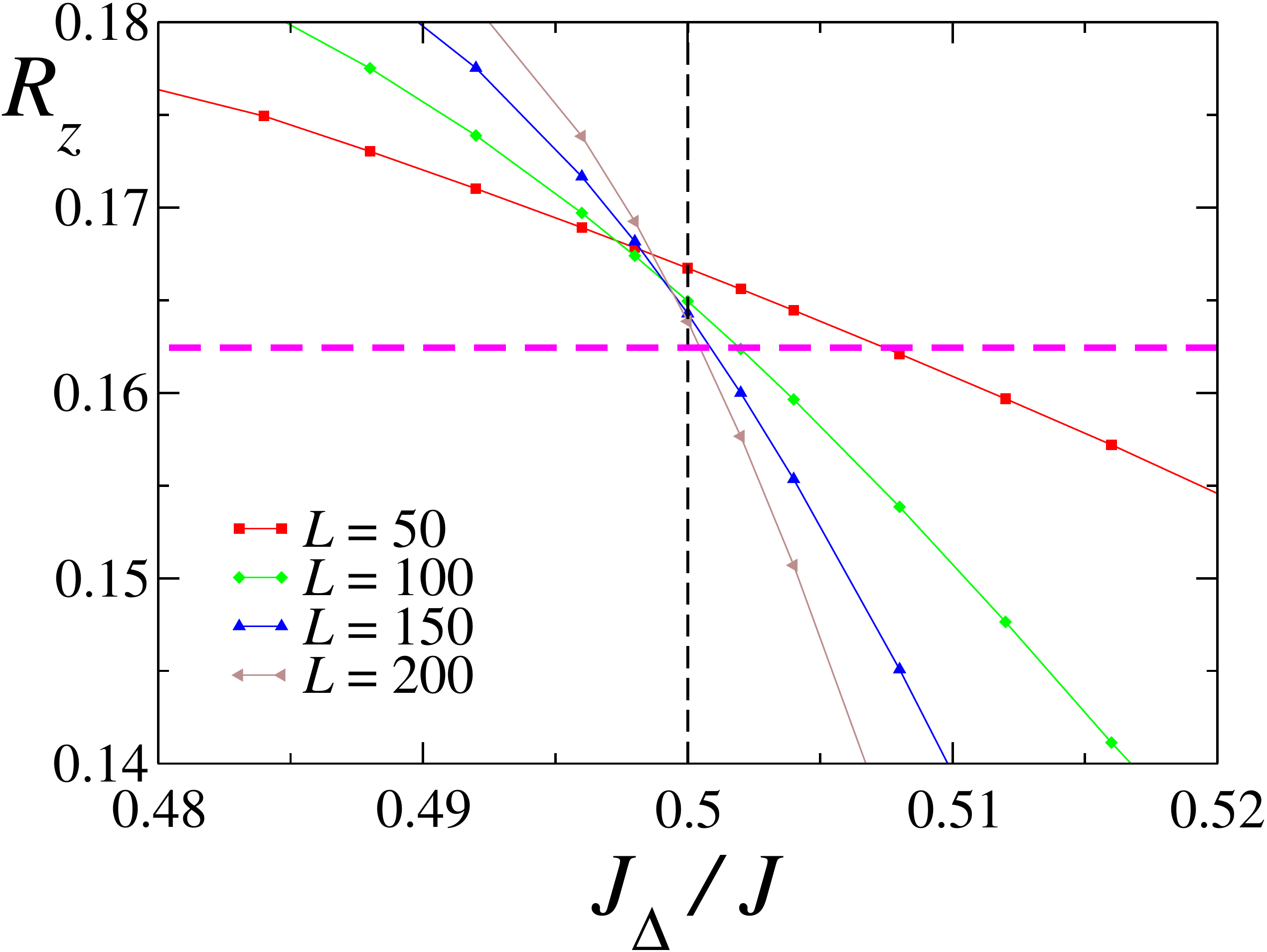} \includegraphics[width=0.5\linewidth]{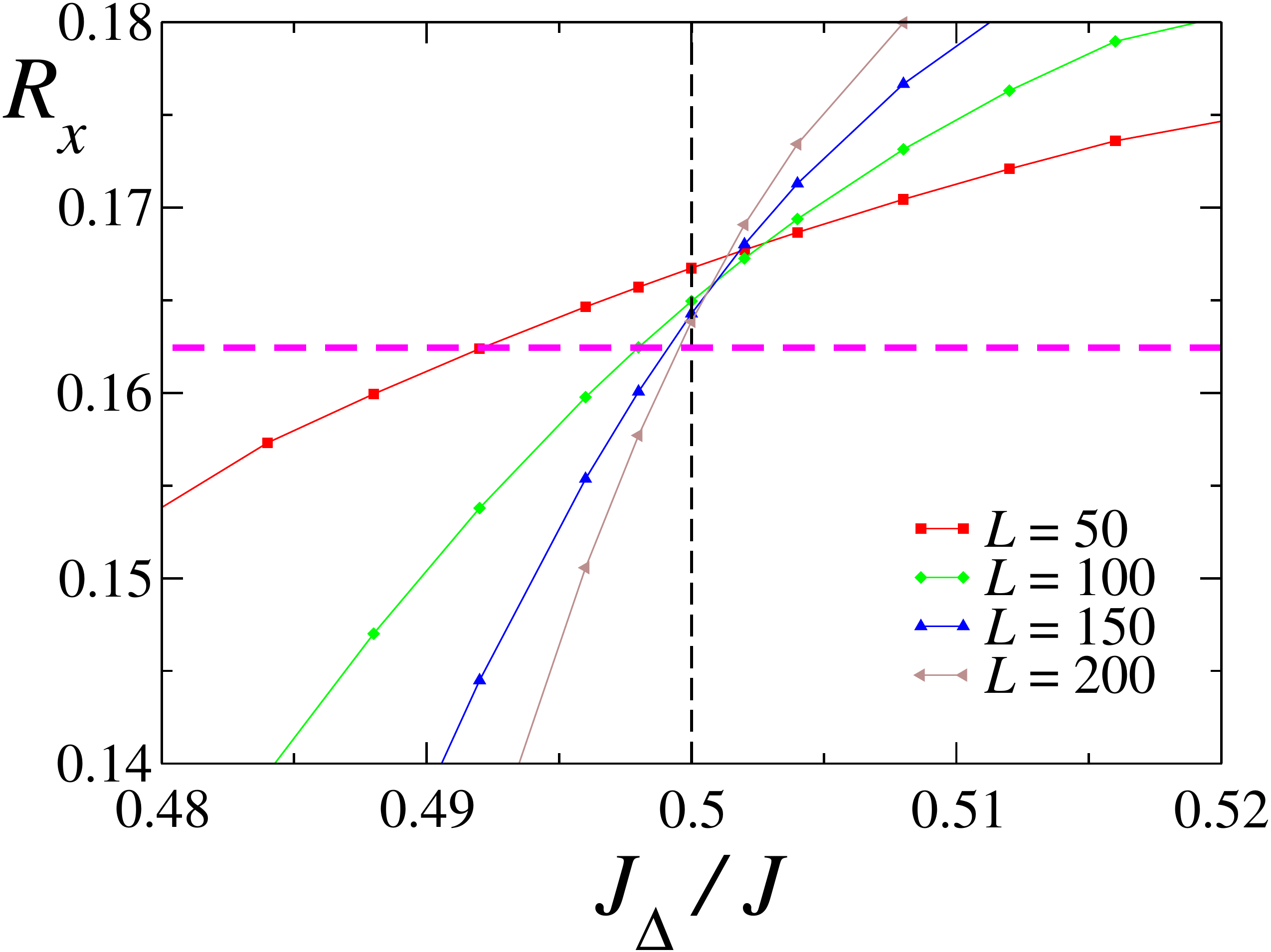}
  \caption{(color online).  Finite-size scaling study of the
    transition between the two ferromagnetic phases in the case
    $\varphi = 0$.  $R_{z} = \xi_{z}/L$ (left) and $R_{x} = \xi_{x}/L$
    (right) are plotted as functions of the couplings $J_{\Delta}/J$
    for $J_z = -3\, J$.  Curves corresponding to different system
    sizes cross at the critical point, which is unique and exactly
    known,~$(J_{\Delta}/J)_c = 1/2$. The dashed line shows the predicted
    value of $R^\star = 0.162445...$ in the $L\to \infty$ limit.      }
  \label{fig:phi0} 
\end{figure}

To begin with, we show results at $\varphi=0$, i.e. for the plain XYZ model,
at $J_z/J = -3$. In this case we expect a single transition point
at $J_\Delta/J=1/2$ between ferromagnetic phases along $x$ and $z$; 
at the transition point one must recover the critical
properties of the XXZ model.  In Fig.~\ref{fig:phi0} we plot the
quantities $R_x$ and $R_z$. Both of them clearly show a crossing
point, which approach the same critical point as expected. Moreover,
at the crossing point the values of $R_x$ and $R_z$ approach the value
$R_x^\star=R_z^\star = 0.162445 \ldots$, as predicted  by 
computations using conformal field theory.

We now move to the case $\varphi \neq 0$, for which no symmetry forces
the XXZ universality class, and thus the nature of the transition 
between the two ferromagnetic phases may drastically change. We focus in
particular on the behavior along the ``cut 3'' of the phase diagram in
Fig.~\ref{fig:phase_diagram_XYZ-DM}, along which $J_z/J = -2$.
We present DMRG results up to $L=400$, see~\ref{app:error}
for technical details on the accuracy of the method.

\begin{figure}
\includegraphics[width=0.5\columnwidth]{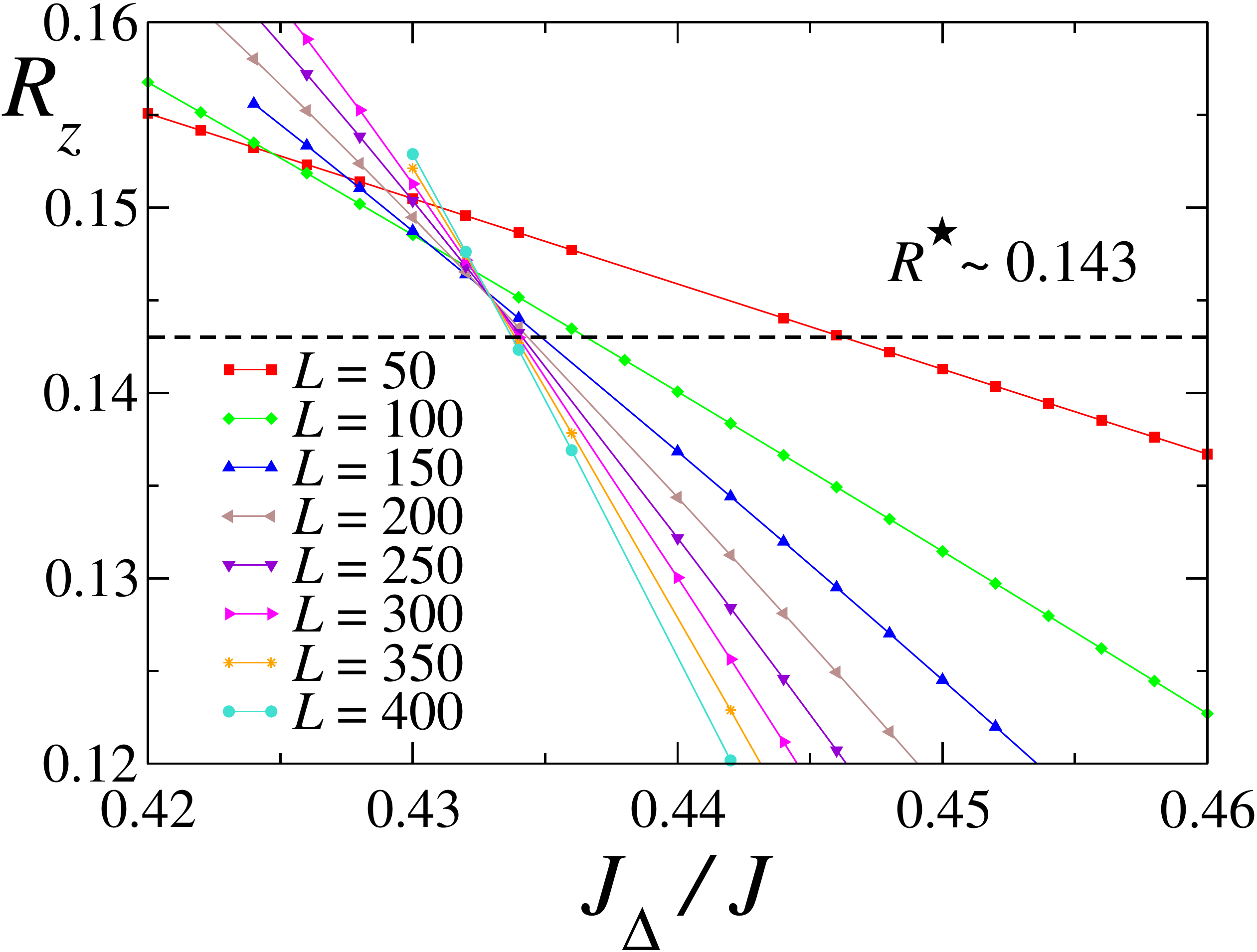} \includegraphics[width=0.5\linewidth]{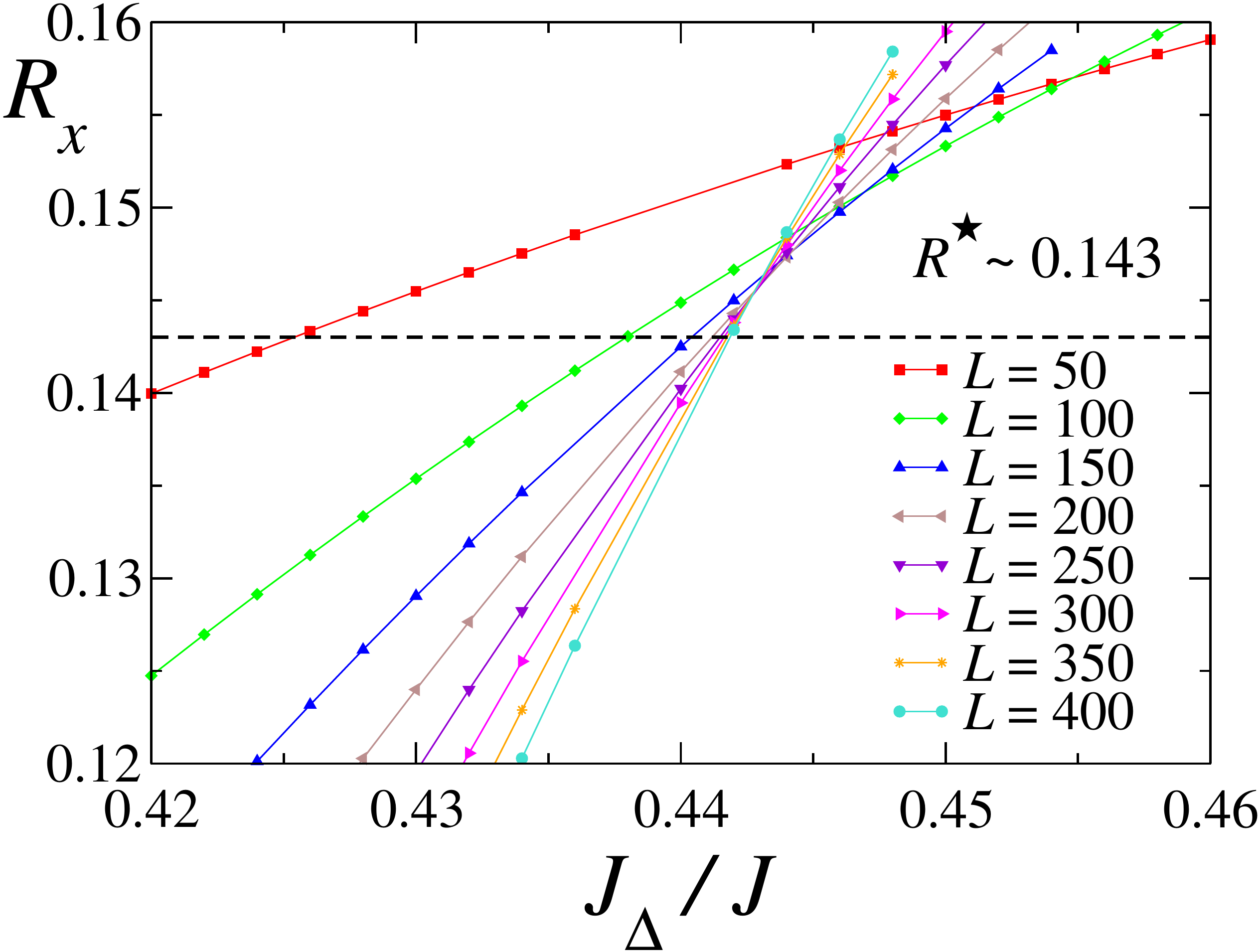}
  \caption{(color online).
    Finite-size scaling study of the transition between the 
    two ferromagnetic phases in Fig.~\ref{fig:phase_diagram_XYZ-DM}  along cut ``3''. The functions
     $R_{z} = \xi_{z}/L$ (left) and
    $R_{x} = \xi_{x}/L$ (right) defined in the main text are plotted as functions of the couplings 
    $J_{\Delta}/J$ for $J_z = -2 J$. 
    Curves for different system sizes show the appearance of
    a crossing point,
    and the plots seem to suggest that there are two distinct ones.}
  \label{fig:xzFM-FSS-1} 
\end{figure}

\begin{figure}[t]
\begin{center}
\includegraphics[width=0.495\textwidth]{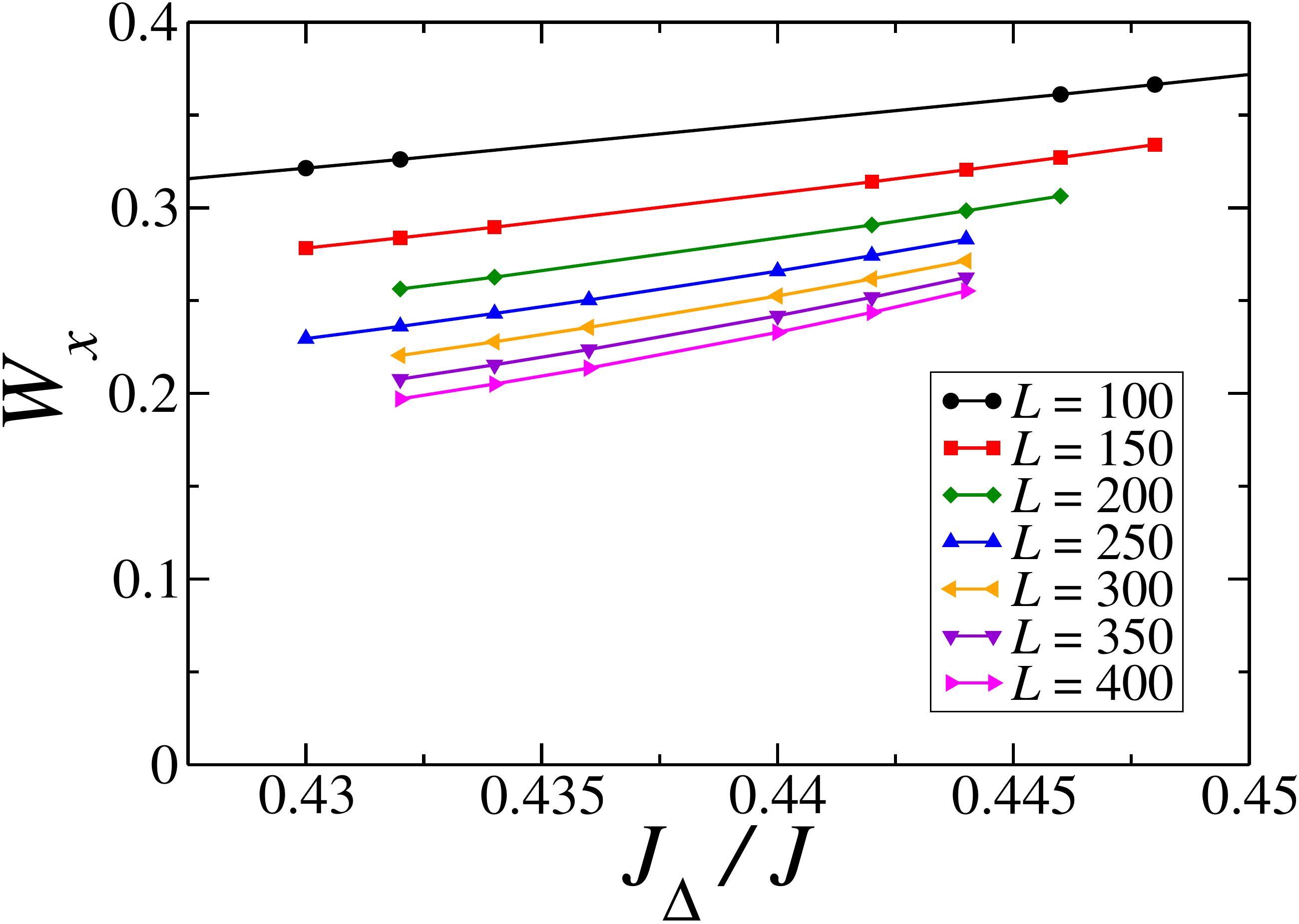}
\end{center}
\caption{(color online). 
The squared magnetization $W_{x}$ as a function
of $J_{\Delta}/J$ for $J_z/J = -2$ (cut ``3'') for several system lengths $L$. 
}
\label{fig:squared:magn}
\end{figure}

Figs.~\ref{fig:xzFM-FSS-1} and \ref{fig:squared:magn} show results for
the ratios $R_\alpha$, see Eq.~(\ref{eq:xixz}), and the
susceptibility-like quantities
\begin{equation}
W_{\alpha} \equiv {1\over L} \sqrt{ \bigg\langle
  \Big(\sum_i \hat S_i^\alpha \Big)^2 \bigg\rangle},  
\label{walpha}
\end{equation}
respectively. These quantities do not show abrupt changes which may
hint at first-order transitions, like those appearing in
Fig.~\ref{fig:LL-xFM}. Therefore, we are lead to exclude first-order
transitions between the $x$-FM and $z$-FM phases.

Within the scenario based on continuous phase transitions, two
possibilities can be envisioned: (i) the transition splits into two
separated critical lines, (ii) the transition remains unique.  For the
case (i), two subcases are possible: (ia) the region in-between
possesses both magnetic orders along $x$ and $z$, (ib) the region
in-between does not possess any magnetic order along $x$ or $z$.

The results  in
Fig.~\ref{fig:xzFM-FSS-1} suggest two distinct transitions.  Indeed
their data sets for different $L$ appear to cluster at different
points, i.e. $J_\Delta/J\approx 0.435$ for $R_z$ and
$J_\Delta/L\approx 0.440$ for $R_x$. Between these two crossing points
the data clearly decrease with increasing $L$, thus suggesting a
disordered phase for both $x$ and $z$ order parameters.  This behavior
seems to exclude the presence of a ordered phase with
both $x$ and $z$ magnetic order.

\begin{figure}[t]
\begin{center}
 \includegraphics[width=0.485\linewidth]{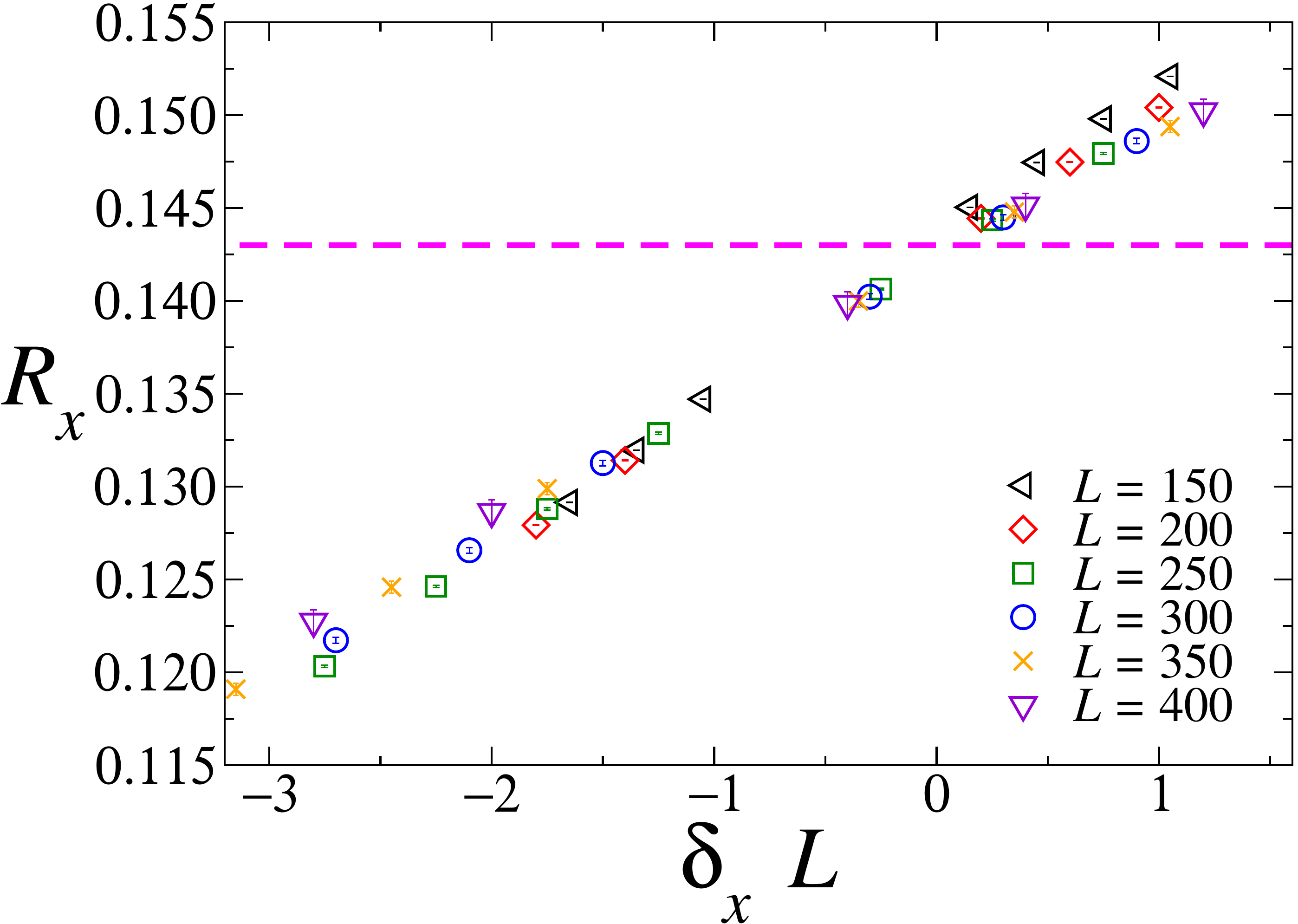}
 \includegraphics[width=0.49\textwidth]{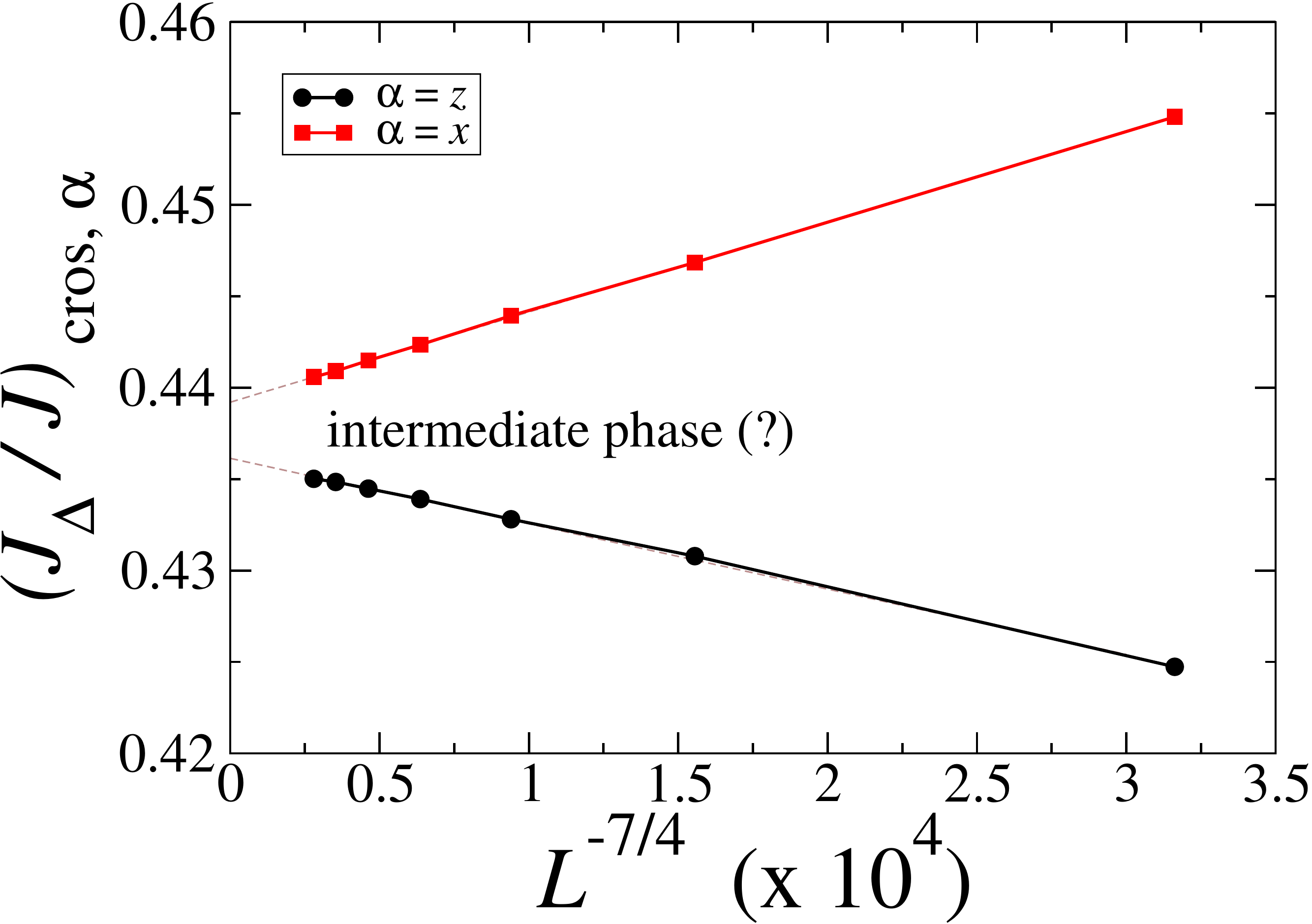}
 \end{center}
 \caption{(color online). 
 Left:
 The quantity $R_x$ is plotted 
 as a function of $\delta_x \cdot L$, 
 using the fitted value (\ref{estx}) for  $(J_\Delta/J)_{c,x}$. 
    We display only data for $L$ in the interval $150 \leq L \leq 400$, 
    for the most accurate DMRG simulations with $m=100$ states. 
    Error bars, estimated as $\Delta R \approx 2 \cdot 10^{-5} (L/200)^5$, are shown (and are of the order or less of the marker size). 
    The dashed line indicates the estimated value $R_\alpha^\star=0.143(2)$.
    An analogous plot is obtained for $\alpha = z$.
 Right: Crossings points 
 $(J_\Delta/J)_{\mathrm{cros},\alpha,L}$ of the 
 $R_\alpha(J_\Delta/J)$ curves
 for lengths $L$ and $L+50$; the cut ``3'',
 $J_z/J = -2$, is considered.
 The scaling behavior $L^{-7/4}$ is highlighted
 and upon extrapolation for $L \to \infty$ (thin dashed line)
 corroborates the possibility that an intermediate phase appears.
 }
 \label{fig:xzFM-FSS-2} 
 \label{fig:crossings}
\end{figure}

Within the scenario of two transitions, with an intermediate disordered
phase for both $x$ and $z$ magnetic variables, a natural hypothesis is
that the two distinct transitions belong to the two-dimensional (2D) Ising
universality class, with $\mathbb Z_2$-like order parameters related
to expectation values of the spin operators
$S_i^x$ and $S_i^z$.
We recall that the critical exponent of the 2D Ising universality class
are $\nu=1$ (length-scale exponent) and $\eta=1/4$
(related to the behavior of the two-point function at criticality).
In order to check this scenario we perform a finite-size
scaling analysis of the data. Assuming a transition in the 2D Ising
universality class, we expect that
\begin{equation}
R_\alpha(J_\Delta/J,L) \approx  {\cal R}(\delta L);
\label{rstar}
\end{equation}
where 
$\delta\equiv  
J_\Delta/J -(J_\Delta/J)_{\alpha,c}$
and
${\cal R}(w)$ is a universal
function (apart from a trivial normalization of the argument),
generally depending on the boundary conditions.
Corrections to the above scaling behavior are suppressed by powers of
$1/L$, in particular the leading ones are~\cite{PV-14} $O(L^{-3/4})$.
In order to determine the critical values of $J_\Delta/J$ for the two
transitions, we fit the data around the crossing point using the
simple ansatz
\begin{equation}
R_\alpha(J_\Delta/J,L) =  R^* + c\, \delta_\alpha L,
\label{ansatz}
\end{equation}
where we keep only the
  first order of the expansion of the r.h.s.  of Eq.~(\ref{rstar}),
  which should provide a good approximation sufficiently close to the
  crossing point. The data in the fit are selected using
  self-consistent scaling conditions with increasing
  $L$~\cite{Vicari:2012}: we select those satisfying
  $-\varepsilon_1<R_\alpha/R^\star-1\lesssim \varepsilon_2$ with
  $\varepsilon_1 \ll \varepsilon_2\approx 0.1$ (the asymmetry between
  $\varepsilon_{1,2}$ is essentially 
  due to the fact that
  the data in the ordered phase are expected to be
  be less contaminated by
  the other degrees of freedom).  The fit neglects the $O(L^{-3/4})$
  corrections.  Their effect is  kept under control by checking the
  stability of the fit results for different values of $L_{\rm min}$,
  which is the minimum system size we have considered.  This numerical
  analysis shows that the data are consistent with the 
  hypothesis of two Ising
  transitions, yielding the estimates
\begin{equation}
(J_\Delta/J)_{x,c} = 0.441(2),\qquad R_x^\star=0.143(2),
\label{estx}
\end{equation}
and 
\begin{equation}
(J_\Delta/J)_{z,c}=0.435(2),\qquad  R_z^\star=0.143(2),
\label{estz}
\end{equation}
where the errors are such to take into account the variation of the
results varying $L_{\rm min}$.  In the fits we take also into
account the precision of the data, which is estimated to be roughly
$\Delta R \approx 2 \times 10^{-5} (L/200)^5$, see
\ref{app:error}.  The quality of this analysis is demonstrated 
by Fig.~\ref{fig:xzFM-FSS-2}, where 
the data of  $R_x$ versus the scaling variable $\delta_x L$ (with $\delta_x =
J_\Delta/J-0.441$) show a good collapse with
increasing $L$. Analogous results are obtained for 
$R_z$ at the other transition.

As a further check of this scenario, in Fig.~\ref{fig:crossings} we
plot the crossing points of the data of $R_x$ and $R_z$ for different
chain lengths.
General finite-size scaling arguments predict that they must
converge to the critical point.  In the case of the Ising universality
class, the renormalization-group analysis of Ref.~\cite{PV-14}
predicts that the crossing points of the ratio $R_\alpha$ must
converge to the critical point with $O(L^{-7/4})$ corrections.  The
data in Fig.~\ref{fig:crossings} nicely support this behavior. They
appear to extrapolate to two different critical points, in agreement
with the estimates reported in Eqs.~(\ref{estx}) and (\ref{estz}).

However, we should also mention an apparent contradiction 
 with the hypothesis of Ising transitions.  For the Ising
universality class with open boundary conditions the value of
$R^\star$ can be computed exactly~\cite{PV-14}, obtaining
 $R^{\star} =
0.159622 ...$.  But this is not compatible with the estimates of
$R_x^\star$ and $R_z^\star$, cf. Eqs.~(\ref{estx}) and (\ref{estz}).  
We ascribe this inconsistency to the residual effects
of the DM interactions,
which may somehow induce nontrivial effective boundary conditions for the
Ising critical modes, 
and thus be responsible for the
mentioned discrepancy.

Concluding, these analyses provide evidence of the 
presence of a new disordered phase in a narrow region of the phase diagram
between the $x$-FM and $z$-FM phases, see Fig.~\ref{fig:phase_diagram_XYZ-DM}.  
The critical behaviors at the two transitions appear
overall consistent with two Ising transitions. However, 
these results should not be considered as a
conclusive analysis of the problem.  Since the two transitions are
very close, we cannot exclude that we are just observing a crossover,
and that the two distinct crossing points will eventually converge
towards a unique critical point for larger values of $L$, as in
scenario (ii).  In this respect, DMRG simulations for significantly
larger system sizes are required to definitely exclude such a
possibility. 
We have however presented strong evidence that this should not be the case.

\section{Spin-Orbit-Coupled Bosons and Fermionic Nanowires}
\label{sec:SOCbosons}

Finally, let us discuss the implications of the previous findings 
for the two specific models of bosons and fermions 
introduced in Section~\ref{subsec:SOCbosons}.

\subsection{Bosons}

We are interested in a strongly-interacting lattice model,
and the system has been driven to a MI; 
we investigate its magnetic properties in the 
experimentally-relevant case of 
$g \sim 0$, $\alpha \in [0, \pi/4]$.
The restriction on $\alpha$ descends from 
(i) the restriction on $\varphi$ previously discussed,
$\varphi \in [0, \pi/2]$, and (ii) Eqs.~(\ref{eq:BH:SOC}),
which imply $\alpha = \varphi/2 + m \pi$, $m \in \mathbb Z$.

The case $g=0$ is particularly simple, 
as it implies $J_{\Delta}=0$ and $J_z = -2J$ independently of $\alpha$.
Thus, as discussed in Sec.~\ref{subsec:exact}, when the interaction is isotropic in spin space, the 
critical properties of the system are those of a ferromagnetic Heisenberg model.
The independence of such properties on $\alpha$ is another way
of stating that for isotropic interactions spin-orbit coupling can be gauged away.

For $g\neq 0$, the manipulation of Eqs.~(\ref{eq:BH:SOC}) shows that
only a subregion of the plane $(J_z/J, J_{\Delta}/J)$ in Fig.~\ref{fig:phase_diagram_XYZ-DM} is accessible:
\begin{equation}
 \frac{J_\Delta}{J} = \pm \frac 12 \cos (\varphi) \, 
  \left( \frac{J_z}{J} +2 \right).
  \label{eq:bosonicPD}
\end{equation}
If we consider the case $\varphi =1$ studied in 
Fig.~\ref{fig:phase_diagram_XYZ-DM}, the system explores
only the phase with ferromagnetic order along $z$ and the LL phase.
Extrapolating the fact that the disordered phase appears for every $\varphi$
and maintains a slope $-1/2$ in the  $(J_z/J, J_{\Delta}/J)$ plane, we conclude that 
a SOC bosonic MI cannot enter the ferromagnetic phase aligned along $x$
for $J_z/J < -2$ and for any value of $\alpha$.
On the other hand, for $\varphi = 0$ the system 
is mapped onto a XZZ model and thus explores the 
phase with ferromagnetic order along $x$ for $J_z/J > -2$~\cite{Lukin_2003}.
For continuity, this may extend to $\varphi \gtrsim 0$.

Let us finally mention the special value 
$\alpha = \pi/4$, for which $J_{\Delta}=0$ 
and the system is mapped onto the XXZ model, 
which does not entail any ferromagnetic phase aligned along $x$.
In particular, for this case:
\begin{equation}
 \frac{J_z} J = -2 \frac{1+ g}{1-g},
 \quad
 e^{i \varphi} = i.
\end{equation}
Thus, for $g \gtrsim 0$ the system enters a gapped ferromagnetic phase, 
whereas for $g \lesssim 0$ the phase is a critical LL.

\subsection{Fermions}

Let us now briefly comment on the implications 
of the phase diagram in Fig.~\ref{fig:phase_diagram_XYZ-DM} 
on the topological properties of the fermionic model in Eq.~(\ref{p-wave}).
The topological phase with Majorana edge modes 
corresponds to the ferromagnetic ordered phase oriented along $x$.
In the absence of interactions, $J_z = 0$, a finite supercurrent 
$\varphi \neq 0$ diminishes its extension to the advantage
of the gapless LL region, which thus cannot be topological.
In the opposite case of strong attractive interactions, 
$J_z /J \to -\infty$,
the model~(\ref{p-wave}) 
corresponds to a simple model of attractive fermions
without topological properties; we can thus conclude that
the ferromagnetic phase along $z$ is devoid of 
protected edge modes.
Even if there is an appropriate Jordan-Wigner transformation 
that maps the $z$ ferromagnetic spin phase 
to a fermionic system with Majorana modes, 
it does not coincide with the mapping used 
for deriving the Hamiltonian~(\ref{p-wave}).
From the phase diagram we can see that a finite attractive interaction increases the critical current that is required to destroy the topological phase. 
It is rather intriguing to investigate what may be the fermionic 
properties of the disordered phase appearing in between the ferromagnetic phases
and to assess whether it is devoid of topological properties.
We leave this analysis for future work.

\section{Conclusions}
\label{conclusions}

In summary, we have analyzed the XYZ spin-$1/2$ chain in presence of 
Dzyaloshinsky-Moriya interactions. It presents a rich phase diagram, 
depicted  in Fig.~\ref{fig:phase_diagram_XYZ-DM},
which has been thoroughly studied in its ferromagnetic region, which is 
most relevant for spin-orbit-coupled bosonic gases loaded in 1D optical lattices.
First-order quantum phase transitions separate 
gapless Luttinger-liquid phases from gapped ferromagnetic phases.
Depending on the relative strength of the couplings,  
such ferromagnetic order can develop along different axes.
The study of the direct phase transition between two
ferromagnetic phases has proven to be particularly intriguing.
Indeed, the Dzyaloshinsky-Moriya term 
breaks the symmetry that in the XYZ model
forces that transition to be unique 
and of the XX universality class.
Our investigation suggests that such critical line
may split into two Ising-like phase transitions,
which are characterized by means of a finite-size scaling analysis of the correlation length.

Our results are relevant for the characterization 
of the phase diagram of one-dimensional bosons 
in optical lattices in the presence 
of spin-orbit coupling and anisotropic spin interactions.
Moreover, they allow the quantitative assessment 
of the stability of the topological phase of the Kitaev chain
characterized by two zero-energy Majorana edge modes
in presence both of interactions and of an external current.
It is fascinating to speculate an extension of this study
to ladder geometries, where additional degrees of freedom
may give rise to new exotic phases~\cite{Sun_2013}.

During the completion of this manuscript we became aware of three works where
one-dimensional lattice bosons with spin-orbit coupling are studied by means of 
density-matrix renormalization-group algorithms~\cite{Zhao:2014, Piraud:2014, Xu:2014}.

\section*{Acknowledgments}
We thank P. Calabrese for fruitful discussions. 
This work was supported by the EU IP-SIQS (Grant agreement No 600645), and the Italian MIUR through PRIN (Project
2010LLKJBX) and through FIRB (Project RBFR12NLNA). 
L. M. is supported by Regione Toscana POR FSE 2007-2013.
S. P. is supported by DOE under Grant No. DE-FG02-05ER46204.

\appendix

\section{Error Estimate for DMRG Data}\label{app:error}

\begin{figure}[t]
\begin{center}
 \includegraphics[width=0.4875\textwidth]{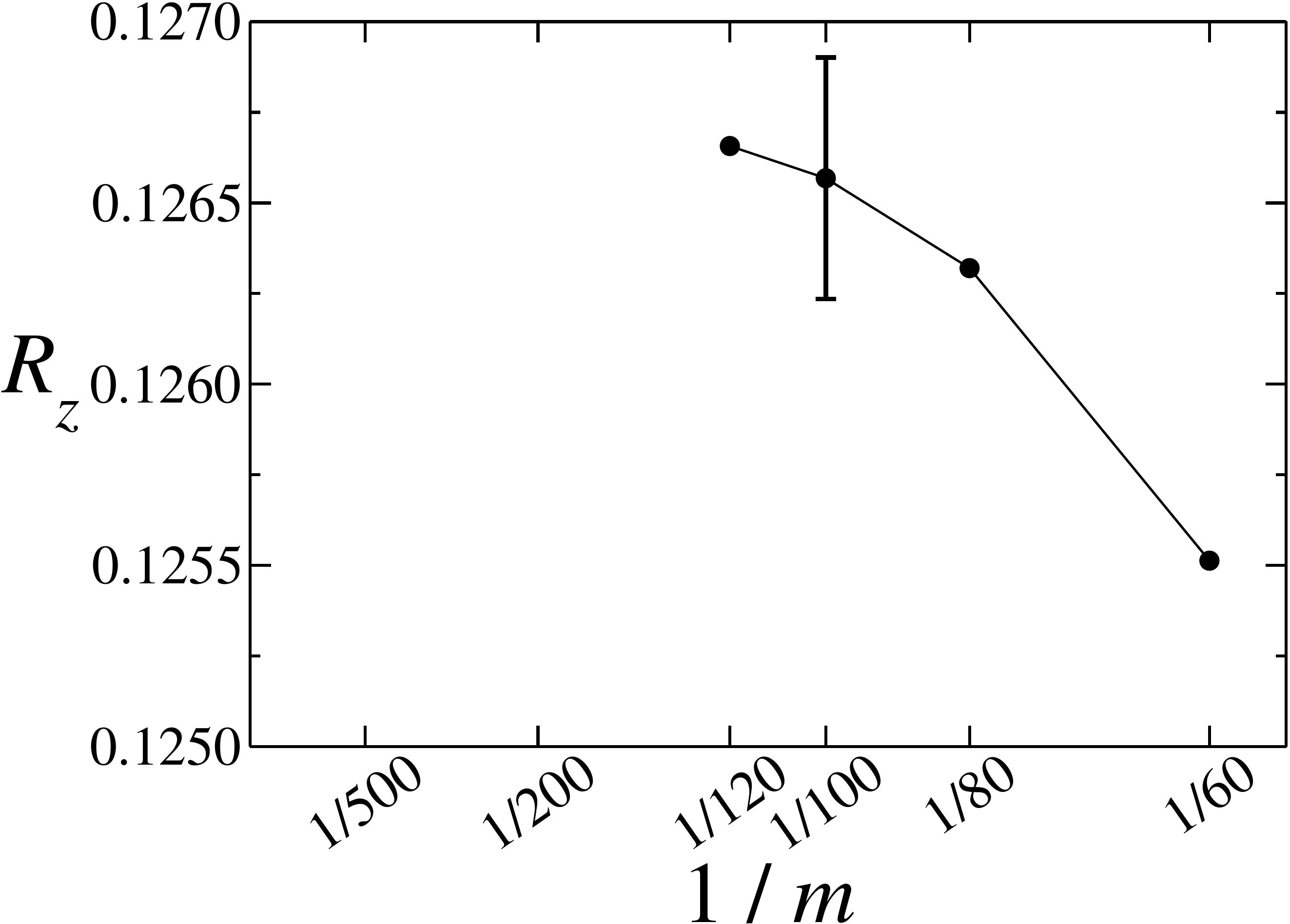}
 \includegraphics[width=0.5025\textwidth]{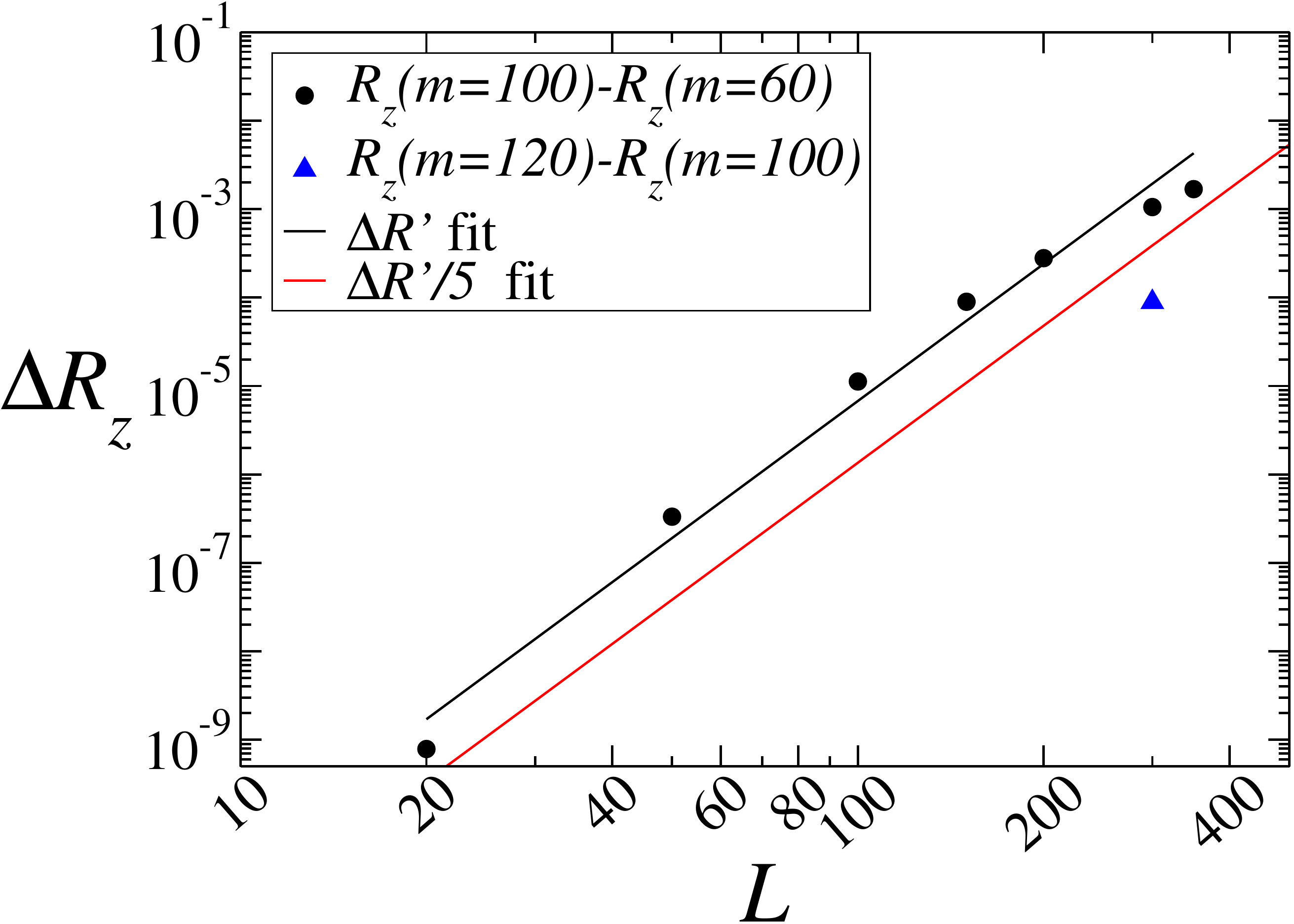}
\end{center}
\caption{(color online).
Accuracy of DMRG simulations for different values of $m$.
(Left) $R_z$ computed 
at $J_\Delta/J = 0.434$, $J_z/J = -2$, $\varphi = 1$
for a system of length $L=300$.
The number of kept states is $60$, $80$, $100$ and $120$;
a clear convergence behavior appears.
(Right) The error is estimated comparing outcomes
for different $m$. The differences
$R(m=100)-R(m=60)$
and $R(m=120)-R(m=100)$   are shown.
The former is fitted by $\Delta R' = 10^{-4} (L/200)^5$.
The error bars on the data at $m=100$
are heuristically estimated considering 
$\Delta R \equiv \Delta R'/5$ (see also the bars in the left plot).
}\label{fig:app:error}
\end{figure}

The accuracy parameter for a DMRG calculation, which characterizes
the outcome of the simulation, $|\Psi_{\rm DMRG} \rangle $, 
is the so called ``number
of kept states $m$'', that is the effective maximal Hilbert space dimension
of each block~\cite{Schollwock:2011, Dummies}. Such number indicates 
that its Schmidt decomposition entails at most $m$ states.
Clearly, the larger is $m$ the better a target state $|\Psi\rangle$
can be approximated.
In Fig.~\ref{fig:app:error} we show the different values of $R_z$
obtained for different values of $m$ for a specific point of
the phase diagram.
In order to compute the error for the data at $m=100$,
which are used in the finite-size-scaling in the text,
we performed some simulations at $m = 120$,
which were however computationally too
demanding for a complete program of simulations
(note also that the model does not conserve any magnetization,
i.e. a symmetry that significantly lowers technical intricacies).
It is possible to observe that
$R(m=120)-R(m=100)\sim(R(m=100)-R(m=60))/5$.
The error formula proposed in the text,
$\Delta R \approx 2 \cdot 10^{-5} (L/200)^5$,
is obtained via a fit
of $R(m=100)-R(m=60)$ close to the quantum phase transition
(see Fig.~\ref{fig:app:error}).
A quantitative improvement of our finite-size-scaling analysis
requires the study of significantly larger systems, 
for which numerical difficulties increase exponentially.

\end{document}